\documentclass[%
 reprint,
 amsmath,amssymb,
 aps,
]{revtex4-1}
\usepackage{natbib}
\usepackage{graphicx}% Include figure files
\usepackage{dcolumn}% Align table columns on decimal point
\usepackage{bm}% bold math
\usepackage[utf8]{inputenc}
\usepackage{comment}
\usepackage{braket}
\usepackage{xcolor}
\usepackage{amsmath}
\usepackage{float}
\usepackage{graphicx}
\usepackage{subcaption}
\usepackage{mwe}
\usepackage[utf8]{inputenc}
\usepackage[export]{adjustbox}
\usepackage{wrapfig}
\usepackage{caption}
\usepackage{lineno}

\begin{document}

%\linenumbers % Enable numberlines
%\preprint{APS/123-QED}
\newcommand{\of}[1]{\left( #1 \right)}
\newcommand{\sqof}[1]{\left[ #1 \right]}
\newcommand{\abs}[1]{\left| #1 \right|}
\newcommand{\avg}[1]{\left< #1 \right>}
\newcommand{\cuof}[1]{\left \{ #1 \right \} }
\newcommand{\pil}{\frac{\pi}{L}}
\newcommand{\bx}{\mathbf{x}}
\newcommand{\by}{\mathbf{y}}
\newcommand{\bk}{\mathbf{k}}
\newcommand{\bp}{\mathbf{p}}
\newcommand{\bl}{\mathbf{l}}
\newcommand{\bq}{\mathbf{q}}
\newcommand{\bs}{\mathbf{s}}
\newcommand{\psibar}{\overline{\psi}}
\newcommand{\svec}{\overrightarrow{\sigma}}
\newcommand{\dvec}{\overrightarrow{\partial}}
\newcommand{\bA}{\mathbf{A}}
\newcommand{\bdelta}{\mathbf{\delta}}
\newcommand{\bK}{\mathbf{K}}
\newcommand{\bQ}{\mathbf{Q}}
\newcommand{\bG}{\mathbf{G}}
\newcommand{\bw}{\mathbf{w}}
\newcommand{\bL}{\mathbf{L}}
\newcommand{\ohat}{\widehat{O}}
\newcommand{\up}{\uparrow}
\newcommand{\down}{\downarrow}
\newcommand{\MM}{\mathcal{M}}
\newcommand{\tW}{\tilde{W}}
\newcommand{\tX}{\tilde{X}}
\newcommand{\tY}{\tilde{Y}}
\newcommand{\tZ}{\tilde{Z}}
\newcommand{\tOm}{\tilde{\Omega}}
\newcommand{\barA}{\bar{\alpha}}

\title{Theoretical survey of unconventional quantum annealing methods applied to a difficult trial problem}%
\author{Zhijie Tang}
    \email{ztang@mines.edu}
    \affiliation{Department of Physics, Colorado School of Mines, 1500 Illinois St, Golden CO 80401}
\author{Eliot Kapit}
    \email{ekapit@mines.edu}
    \affiliation{Department of Physics, Colorado School of Mines, 1500 Illinois St, Golden CO 80401}

% \date{\today}%

\begin{abstract}
We consider a range of unconventional modifications to Quantum Annealing (QA), applied to an artificial trial problem with continuously tunable difficulty. In this problem, inspired by "transverse field chaos" in larger systems, classical and quantum methods are steered toward a false local minimum. To go from this local minimum to the global minimum, all N spins must flip, making this problem exponentially difficult to solve. We numerically study this problem by using a variety of new methods from the literature: inhomogeneous driving, adding transverse couplers, and other types of coherent oscillations in the transverse field terms (collectively known as RFQA). We show that all of these methods improve the scaling of the time to solution (relative to the standard uniform sweep evolution) in at least some regimes. Comparison of these methods could help identify promising paths towards a demonstrable quantum speedup over classical algorithms in solving some realistic problems with near-term quantum annealing hardware.   
\end{abstract}

\pacs{Valid PACS appear here}% PACS, the Physics and Astronomy
\maketitle 

%\tableofcontents

\section{\label{sec:Introduction}Introduction}

Quantum Annealing (QA)~\cite{Finnila,Kadowaki,Das,Johnson2011,Somma2012,Albash2016} is a promising method to solve optimization problems with noisy quantum hardware, with applications in machine learning, artificial intelligence~\cite{Boixo2014,Gili,Neukart,Venturelli_2018,King2018,Venturelli2019}, and many other topics. The time-dependent Hamiltonian of QA is engineered to encode the solutions of classical optimization problems in its ground state. By initializing the system in the ground state of a trivial driver Hamiltonian and evolving the system sufficiently slowly, QA can find the ground state of the target (classical) problem Hamiltonian. However, it is notoriously difficult to predict the performance of QA for realistic problems. Conclusive proof of a quantum speedup over classical methods for real problems remains elusive, with the possible exception of a frustrated magnet systems~\cite{king2019scaling}, where an empirical scaling advantage over classical path-integral Quantum Monte Carlo (QMC) algorithms was shown. To help address this challenge, in this paper we theoretically survey a range of promising extensions to QA applied to a difficult trial problem, and identify a number of potential routes to a quantum speedup. 

In the setup for QA, the total Hamiltonian in the standard uniform sweep evolution is a combination of a driving Hamiltonian $H_0$ and problem Hamiltonian $H_p$,
\begin{equation}\label{eq:1}
H(t) = (1-s(t))H_0 +s(t)H_p
\end{equation}
with the ground state of $H_0$ being easy to prepare. $H_0$ and $H_p$ do not commute and the time-dependent annealing parameter $s(t)$ controls the time evolution of the system. The standard uniform sweep starts from $s(0) = 0$ and ends with $s(t_f) = 1$. Different functional choices for $s(t)$ may vary the efficiency of finding the ground state, such as a `reverse annealing schedule'~\cite{James}. It's intuitive to see that slowing down the annealing process in the vicinity of minimum gap can help increase the success probability, and this tuning is required to recover the quantum speedup in the QA formulation of Grover's search problem~\cite{Roland}. However, the instantaneous minimum gap value and location is not knowable in most realistic problems, and such fine tuning is often frustrated by noise, so we will study the simplest form, a linear schedule, $s(t)$: $s(t)=\frac{t}{t_f}$ throughout this paper. 

In QA, the ground state of a problem Hamiltonian, $H_p$, encodes the optimization problem solution. Experimentally realistic formulations of quantum annealing are typically arranged to solve quadratic unconstrained binary optimization (QUBO) problems, where the problem Hamiltonian is given by the Ising model, 
\begin{equation}\label{isingproblem}
H_{\mathrm{ising}} = \sum_{<i,j>}J_{ij} \sigma_i^z \sigma_j^z + \sum_{i=1}^{N} h_i \sigma_i^z 
\end{equation}
the ground state of which can be encoded as the solution space of some NP-hard problems~\cite{Lucas}, and given enough additional qubits, any NP complete problem can be expressed in this form. To find the ground state of the problem Hamiltonian, we first prepare the system in the ground state of $H_0$, which is chosen as a uniform transverse field Hamiltonian, 
\begin{equation}\label{eq:3}
H_0 = -\sum_{i=1}^{N} \sigma_i^x.
\end{equation}
The initial ground state is a uniform superposition state in the computational basis. The quantum adiabatic theorem states that as long as the annealing evolution is slow enough, the system remains in the instantaneous eigenstate of the time-dependent Hamiltonian at all times. This theorem also provides a widely used criterion that, with the linear annealing schedule, the total adiabatic evolution time $t_f$ to find the ground state with high probability has an inverse minimum gap squared dependence,
\begin{equation}\label{eq:4}
t_f \propto \frac{W}{\Delta_{min} ^2}.
\end{equation}
Here, $\Delta_{min}$ is the minimum energy gap, $W$ scales as the total energy change of the final ground state $\ket{0}$ over the entire evolution: $W \sim \bra{0}H_0\ket{0}-\bra{0}H_p\ket{0}$. Note that this result is a \emph{worst case} scaling estimate of the time to solution, and a variety of diabatic effects can substantially increase performance--we will encounter a number of examples of this later in this work. 

In cases where the system undergoes a first order transition~\cite{jorg2010energy}, $\Delta_{min}$ typically decreases exponentially with the system size $N$, and the corresponding evolution time $t_f$ (and thus, time to solution) grows exponentially. For hard optimization problems which suffer from such phase transitions, many new schemes have been proposed to accelerate QA, such as the use of non-stoquastic Hamiltonians~\cite{farhi2002quantum,Seki,Elizabeth,Seki_2015,nishimori2017exponential,Hormozi}, inhomogeneous driving of the transverse field~\cite{Susa2018a,susa2018quantum}, and oscillatory transverse fields (RFQA)~\cite{Kapit2017}. We  study a range of examples drawn from these works.

% Besides these methods, we will show that an oscillating transverse field method(RFQA), which includes both non-stoquastic Hamiltonians and stoquastic Hamiltonians can also provide performance enhancement of quantum annealing.

To investigate a number of new methods from the literature, we make our own artificially difficult problem Hamiltonian, partially inspired by previous studies of ``spike problems"~\cite{farhi2002quantum,kong2015performance}, rather than studying QUBO problems directly as was done in~\cite{Elizabeth}. In this problem, which we call the Asymmetric Magnetization Problem (AMP), local searches and QA steer the system toward a false minimum. This `wrong way steering' makes finding the true ground state exponentially difficult. The difficulty exponent associated with the AMP can be continuously tuned for further investigation of the performance of these alternative QA methods with problem hardness. Further, the energy landscape depends only on total magnetization $m$, making it easier to study analytically. We do not consider noise in this paper, as none of the methods we consider require fine tuning. We expect these methods to be resilient to noise described by the empirical noise model for superconducting flux qubits. There is strong theoretical evidence for this resiliency in the case of RFQA~\cite{Kapit2017}, and both theoretical and experimental evidence for inhomogeneous driving~\cite{Berry2009,Susa2018a,susa2018quantum,adame2020}.

The rest of the paper is organized as follows. In Section II, we introduce our trial problem; in Section III, we provide an analytical prediction of its minimum gap. Section IV discusses the performance of the standard uniform sweep applied to this problem, against which we benchmark all other methods.  We then introduce inhomogenous driving, transverse couplers and RFQA, comparing their behavior with the standard uniform sweep routine in Sections V and Section VI separately. The final section summarizes our results and provides comments on the performance of these methods. 
\begin{figure}[!ht]
\centering
  \includegraphics[width=10cm,height=7.5cm]{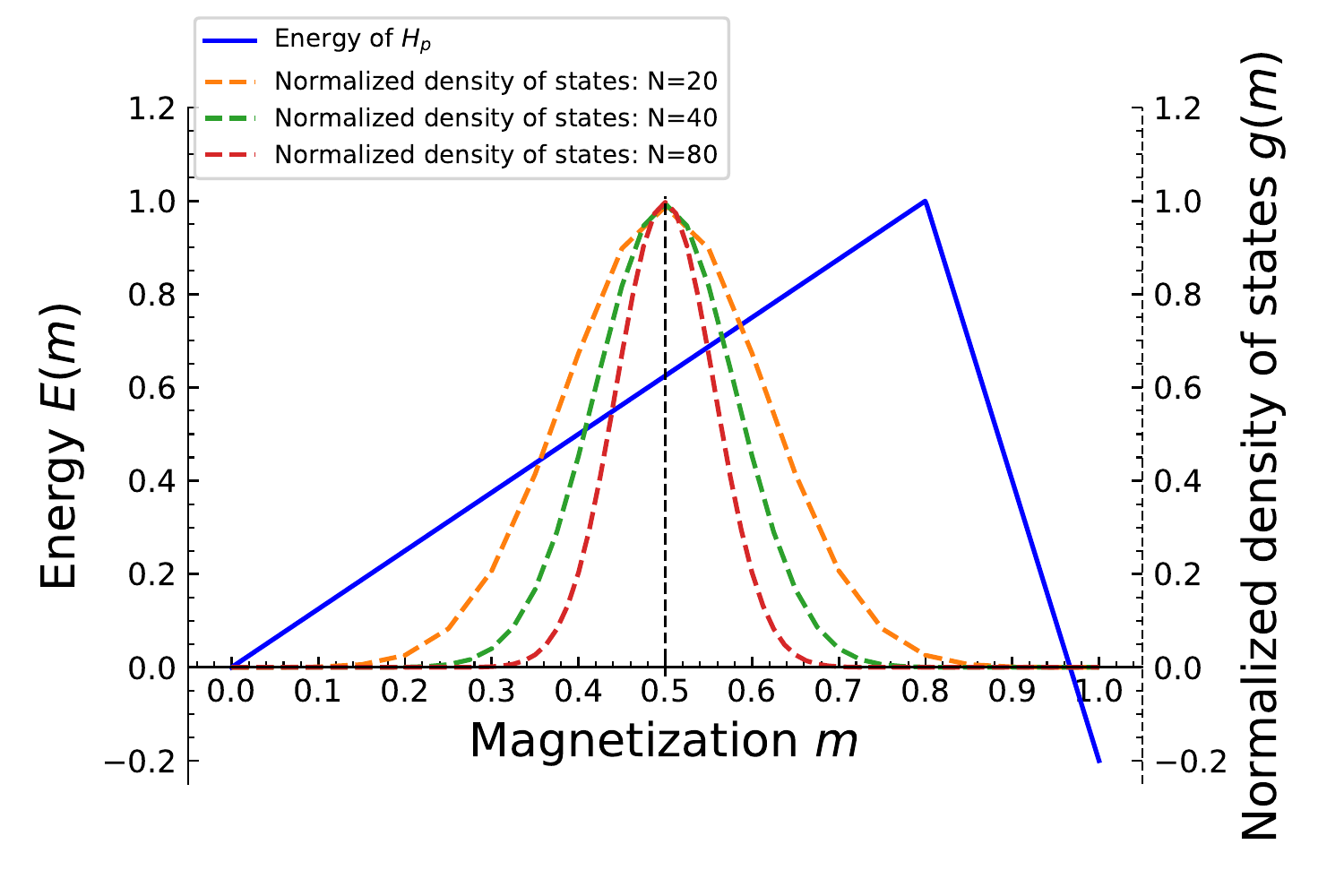}
\caption{Density of states distribution and energy of the problem Hamiltonian in the AMP model. The x axis is the total magnetization $m$, the y axis represents both the energy spectrum of the problem Hamiltonian(left) and the density of states distribution(right). The dashed lines are the density of states for system sizes of N=10, 11, 12. The distributions become narrower with increasing system size. The solid blue line is the energy landscape of $H_p$. In our model, the distribution follows a Gaussian distribution, centered at $m = 0.5$. Comparing the distribution of density of states with the energy spectrum of the problem Hamiltonian, we see the peaks of the density of states are distributed behind the global maximum. The asymmetry of the $H_p$ results in an exponentially difficult problem. }
\label{distribution of state density}
\end{figure}

\section{\label{sec:problem model}Asymmetric Magnetization Problem }
While conventional Ising Hamiltonians can encode nearly any combinatorial optimization problem, we choose an artificial toy problem model to better study a variety of new methods to find its solution. This is in part because extracting exponential difficulty scaling is notoriously difficult and unreliable in random structured problems, at least when $N$ is small enough for exact classical simulation. Our artificial problem, the ``AMP'', has the problem Hamiltonian defined as function of total magnetization $m$,
\begin{equation}\label{eq:5}
  \begin{aligned}
    H_p = f(m); \; \;   m=\frac{1}{2N}\sum_{i=1}^{N}(1+\sigma_i^z)
  \end{aligned}
\end{equation}
where N is the system size, $\sigma_i^z$ is the Pauli matrix with discrete eigenvalues $\pm 1$, $f(m)$ is designed to have two competing ground states at $m=0$ and $m=1$ with all spins down and all spins up. The form of $f(m)$ is controlled by two free parameters, $A$ and $x_p$:
 \begin{equation}\label{eq:6}
  f(x) =
    \begin{cases}
      \frac{x}{x_p} & \text{if $x < x_p$}\\
      1-(1+A)\frac{x-x_p}{1-x_p} & \text{if $x \geq x_p$}\\
    \end{cases}       
\end{equation}
Here, $f(1)$ is the true ground state, and $f(0)$ is the false minimum.
$x_p$ defines the location of the global maximum, and $A$ defines the energy difference of the two competing states: $f(0)-f(1) = A$. By adjusting these two parameters, we can continuously tune the difficulty of the problem. 

The difficulty of finding the global minimum in our model is strongly related to the distribution of the density of states. The density of states follows a Gaussian distribution as shown in FIG.~\ref{distribution of state density}, with the most probable initial state centered at $m=0.5$ where half of the spins are flipped from the true ground state, and a global maximum is distributed at $m = x_p > 0.5$. We see the system has a large tendency to get stuck in the local minimum, since the possible initial state is settled behind the global maximum. The wrong way guidance is the generic failure mechanism for classical and quantum optimization algorithms~\cite{Albash2}, as if local guidance from a random initial state tends to point toward the true solution of a problem it can be solved trivially. But if local guidance points toward false minima, then the problem can quickly become hard, and in more realistic problems at large $N$ there are often exponentially many local minima. Multiqubit tunneling between well-separated minima--exactly the process we simulate here--has been identified as a critical bottleneck in many realistic problems ~\cite{knysh2016}.

In the AMP problem, we set the global maximum at $m=x_p$ where $x_p > 0.5$, with two competing ground states at each end. From the density of states distribution in FIG.~\ref{distribution of state density}, we can tell the system has a large tendency to be steered toward the false minimum. In classical algorithms such as simulated annealing, the system will easily get stuck in the false minimum since the possible initial state is mostly distributed around $m=0.5$. The possibility of finding the global minimum is large if the initial instantaneous state of $H_p$ happens to be guessed beyond the global maximum at a position that $m > x_p$, but if the initial state is located at any $m < x_p$, the possibility of climbing the hill is exponentially small. Cost functions similar to the AMP model have been studied in \cite{farhi2002quantum,kong2015performance}, and classical simulated annealing was shown to be inefficient for solving such problems. We will show that the AMP problem is also exponentially difficult to solve with quantum annealing, and we focus on how various modifications to QA compare with a homogeneous transverse field and uniform sweep (the ``default" quantum annealing method) in solving the AMP problem.
\begin{comment}
when the initial instantaneous state of $H_p$ happens to be distributed beyond the global maximum at a position that $x_p<M<=1$
It's the same case in simulated annealing method, although it provides possibility to overcome the global maximum when the initial state is behind the global maximum, it's still can be exponentially hard to solve.
\end{comment}

When applying QA to the AMP problem, performance is bottlenecked by an exponentially small gap at a first-order transition~\cite{jorg2010energy,Bibes2010,Laumann}. As shown in FIG.~\ref{fig4}, the magnetization is entropically steered toward 0 as the system evolves, and all $N$ spins must simultaneously flip to reach the true ground state. The difficulty scaling of the problem model can be tuned by $A$ and $x_p$; smaller $A$ corresponds to a smaller energy gap between the two competing ground states, which intuitively increases the difficulty level of the problem. Similarly, larger $x_p$ moves the peak further away from the center of the density of states, and the system then has larger tendency to be steered to the false local minimum, which also increases problem difficulty. We make an ensemble of problem models with different $A$ and $x_p$ so that we can investigate the relationship between the performance of different methods with the difficulty of the problem models. We make modifications to the traditional QA method and evaluate their performance by numerically calculating the time to solution, and show that the AMP problem is exponentially difficult to solve with quantum algorithms, but modifications to the traditional QA method can lead to substantial improvements in the scaling of the time to solution. The exponential scaling coefficients for each method are listed in Table I.   

Although this toy model is just a simplified artificial problem without a realistic implementation, as described above, it captures the basic bottleneck of most classical and quantum optimization problems. So any method which accelerates finding a solution in the AMP is likely broadly applicable to more realistic cases.

\section{\label{sec:Materials}Analytical prediction of the minimum gap}
The minimum gap $\Delta_{min}$ determines the worst case difficulty of a problem, so analytically predicting it can help us to better assess the behavior of quantum annealing algorithm. To compute it, we use a modified form of the ``forward approximation" $N$th order perturbation theory employed in~\cite{pietracaprina2016forward,baldwinlaumann2016,baldwinlaumann2017,scardicchio2017perturbation,baldwin2018quantum}. In this approximation, the minimum gap is predicted to be 
\begin{equation}\label{eq:mingap}
\Delta_{min} = N!\frac{\prod_{i=1}^{N} \kappa_i}{\prod_{n=1}^{N-1}U_{n}}
\end{equation}
where $\kappa_i$ is the transverse field strength on each site $i$ (evaluated at the critical point $\kappa_c$), and $U_n^{-1}$ is the average of the inverse of the energy difference to flip $n$ spins from either ground state
\begin{equation}\label{eq:oneoveruk}
\frac{1}{U_n} = \Bigg \langle  \frac{1}{\epsilon_n+\delta_{\epsilon_n} - \epsilon_0 - \delta_{\epsilon_0}}\bigg \rangle.
\end{equation}
Here, the $\epsilon$ terms are the classical energies defined in the problem Hamiltonian and the $\delta$ terms are their perturbative corrections from the transverse field, which act to increase the excitation energies in this case. Including these corrections in the energy denominators (which is effectively a resummation scheme) is vital to obtaining relatively accurate predictions; explicitly, for the AMP
\begin{eqnarray}\label{eq:UAMP}
U_{n} &\simeq& n \of{ \frac{1}{x_p} + 2 \kappa_c^2 x_p }, \; \; \; \cuof{n \leq x_p N }, \\
&\simeq& \of{N-n} \of{ \frac{1+A}{1- x_p} + 2 \kappa_c^2 \frac{1- x_p}{1+A} } \; \; \; \cuof{n > x_p N}. \nonumber
\end{eqnarray}

\begin{figure}[h]
\centering
  \includegraphics[width=9.3cm,height=5.5cm]{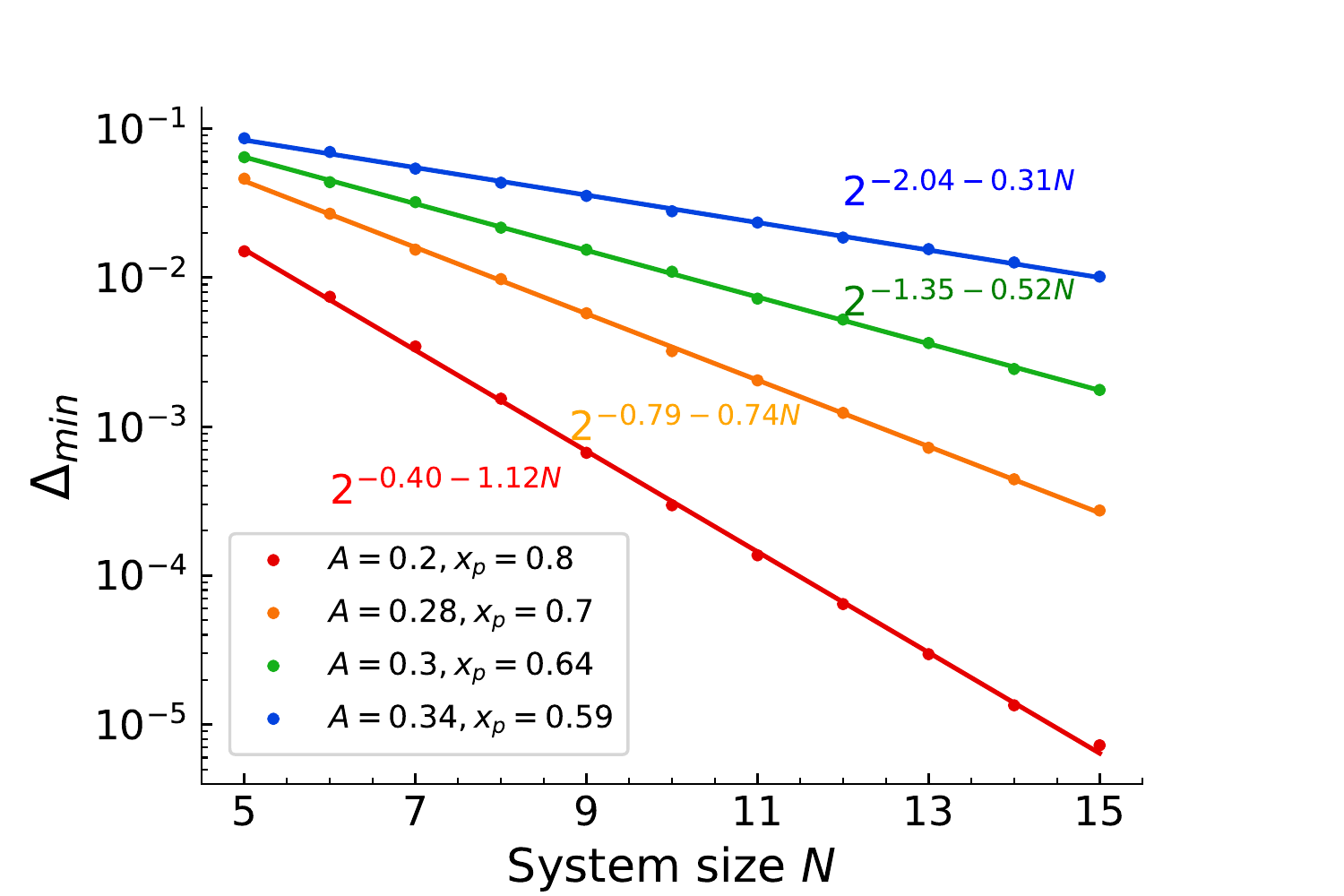}
  \caption{Numerical values of the minimum gap for four parameter sets, with the system size $N$ ranging from 5 to 12 spins. The points are data of numerically estimated minimum gap energies, least squares fits of the minimum gap are red, orange, green and blue solid lines. The minimum gaps in the four problem sets all decrease exponentially with increasing system size.}
  \label{fig2}
\end{figure}

\begin{figure}[h]
\centering
  \includegraphics[width=9.3cm,height=5.5cm]{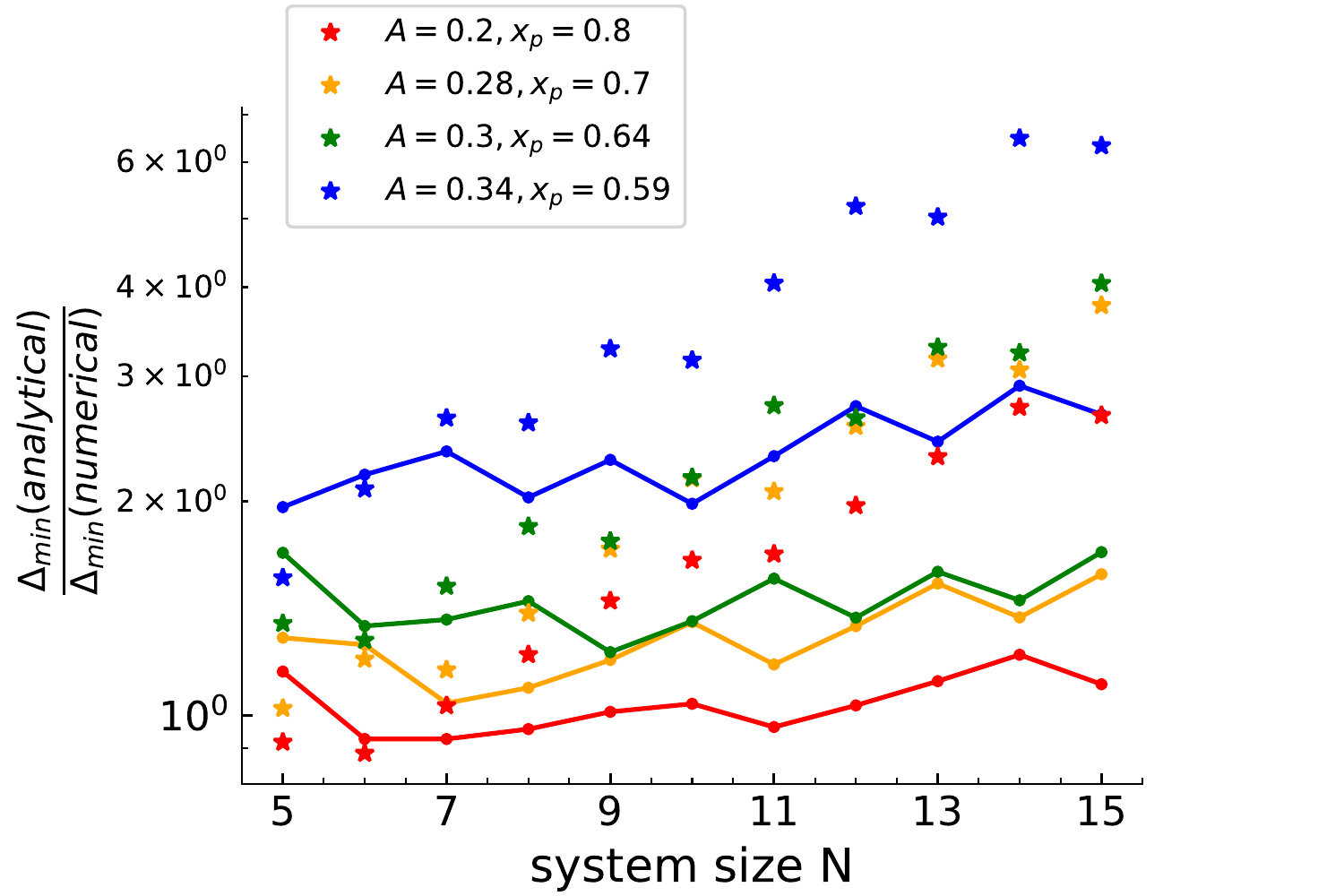}
  \caption{Ratio of analytical and numerical values of minimum gap in four problem sets, system size $N$ ranging from 5 to 12. The analytical values of $\Delta_{min}$ are from the modified forward approximation (Eqs.(\ref{eq:mingap} -- \ref{eq:UAMP})). Star marker represents the ratio of analytical and numerical values of minimum gap, dot-line represents the ratio of analytical and numerical values with a correction term $2\pi/N$. The corrected analytical predictions in the four sets match well with the numerical values.}
  \label{fig3}
\end{figure}

From this expression, it is straightforward to predict the minimum gap in our problem. As shown in FIG.~\ref{fig2}, the exponential fitting of numerical $\Delta_{min}$ as a function of $N$ for descending difficulty problem sets are: $2^{-0.4-1.12N}$, $2^{-0.79-0.74N}$, $2^{-1.35-0.52N}$, $2^{-2.04-0.31N}$. FIG.~\ref{fig3} indicates that the scaling of our theoretical prediction matches well with the numerical result by multiplying by a factor of $2\pi/N$. Eq.(\ref{eq:mingap}) appears to overestimate the true gap by a factor of $\sim N/2\pi$; the reason for this is unclear. Some level of disagreement is expected, however, particularly in the easiest difficulty parameter set. If the coefficient of the problem Hamiltonian is 1, the phase transition for those parameters occurs at $\kappa_c \simeq 1.73$. At such a large value of $\kappa_c$ the ratio of $\kappa$ to the single spin excitation energy approaches unity and thus a perturbative expansion in it may break down.


\section{\label{sec:Procedure}Standard uniform sweep routine}
We first investigate the performance of the standard uniform sweep method, for system sizes $N$ ranging from 5 to 18 spins. In this method, the driving Hamiltonian is a homogeneous transverse field: 
$H_0 = -\sum_{i=1}^{N} \sigma_i^x$. 
The total Hamiltonian is a combination of the drive Hamiltonian and problem Hamiltonian $H_p$
\begin{equation}\label{eq:10}
H(s)=-(1-s)\frac{1}{N}\sum_{i=1}^{N} \sigma_i^{x}+s H_p.
\end{equation}

\begin{figure}[htbp!]
\centering
  \includegraphics[width=9.3cm,height=5.5cm]{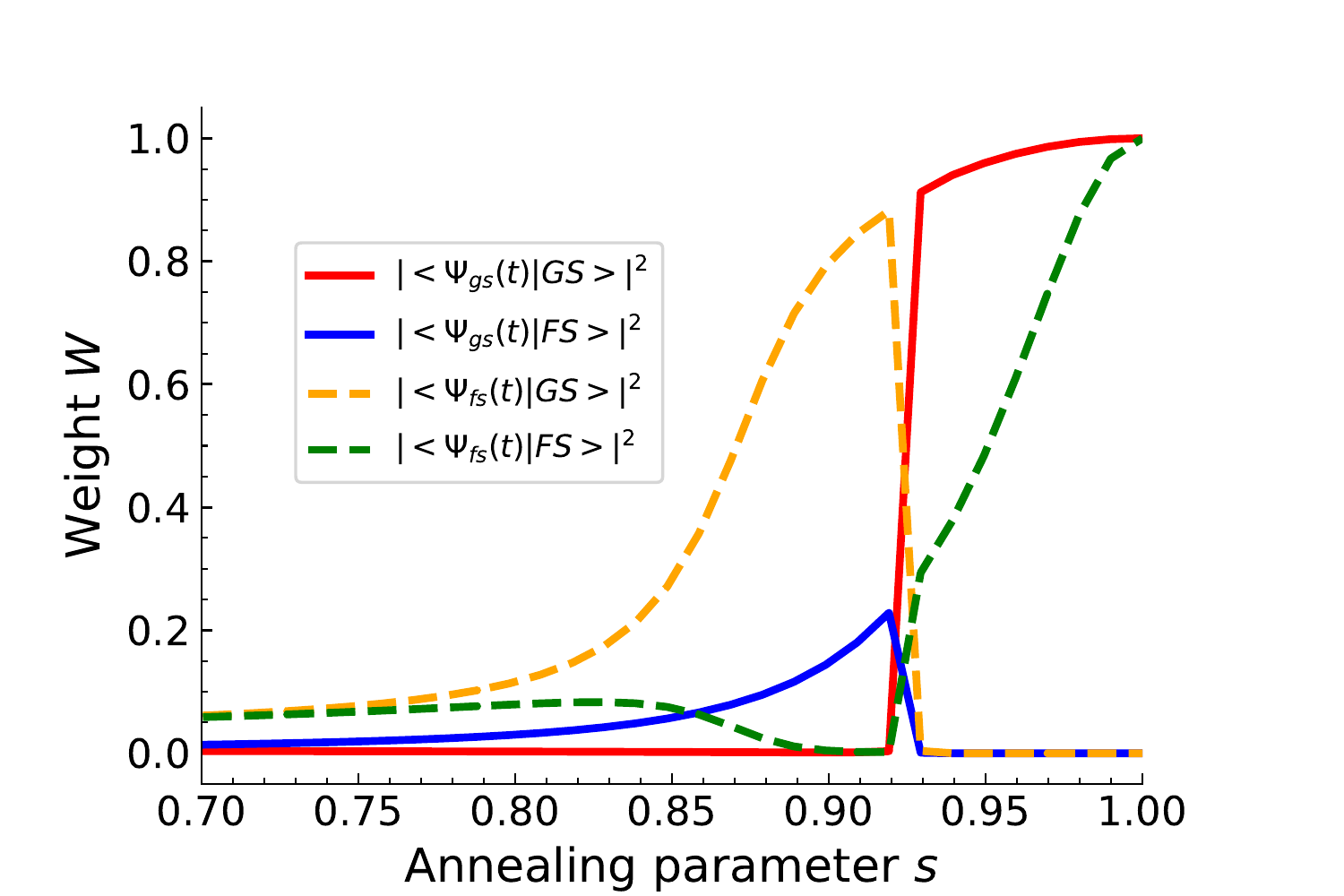}
\caption{The instantaneous overlap of the ground and first excited states with the true and false ground states of the classical problem as a function of the annealing parameter $s$, with system size $N=8$, and the problem Hamiltonian is defined with parameters: $A=0.2,x_p=0.8$. The x axis is the annealing parameter $s$ in a small range from 0.7 to 1, the y axis denotes the probability of getting a specific state. The red and blue solid lines are the overlap of the instantaneous ground state with the true and false ground state of $H_p$, the orange and green dashed lines are the overlap of the instantaneous first excited state with the true and false ground state of $H_p$. The comparison shows that the system is steered toward the false minimum first and all N spins have to flip to reach the true ground state.}
\label{fig4}
\end{figure}
The initial ground state of the system is the ground state of $H_0$, which is a uniform superposition of states corresponding to all possible assignments of bit values with equal weights. As the system evolves, the Hamiltonian linearly interpolates between the transverse field Hamiltonian and the problem Hamiltonian, i.e. starting as $H_0$ and ending in $H_p$. As long as the system stays in the instantaneous ground state, the system will be steered toward the false minimum first, but at some critical $s_c$ the true and false ground states cross and all $N$ spins must flip, as shown in FIG.~\ref{fig4} with red and blue solid lines. For this problem, there is only one avoided crossing in the standard uniform sweep method (though we find multiple crossings when inhomogenous driving is employed). Since the gap at the phase transition point is exponentially small in $N$, unless the evolution is performed extremely slowly, the avoided crossing will be diabatically missed and the probability of finding the true ground state will be suppressed, while the probability of finding the false ground state dominates, as shown in FIG.~\ref{fig4} with orange and green dashed lines.

% At the end of the anneal cycle, the transverse field is turned off, and the quantum fluctuation stops, the system will end with the true ground state of $H_p$. However, the system may transit into the first excited state at some critical point, especially near the minimum gap. We see that the minimum gap can be a bottleneck in our model, small perturbations may excite the system in the vicinity of minimum gap. And when the system stays in the instantaneous first excited state, the probability of finding the true ground state will be suppressed while the probability of finding the false ground state dominates as shown in Fig.4(b). To avoid the excitation, we can either evolve the system sufficiently slow with linear or non-linear annealing schedule or couple the system with cold bath.

\begin{figure}[htbp!]
\centering
  \includegraphics[width=9.3cm,height=5.5cm]{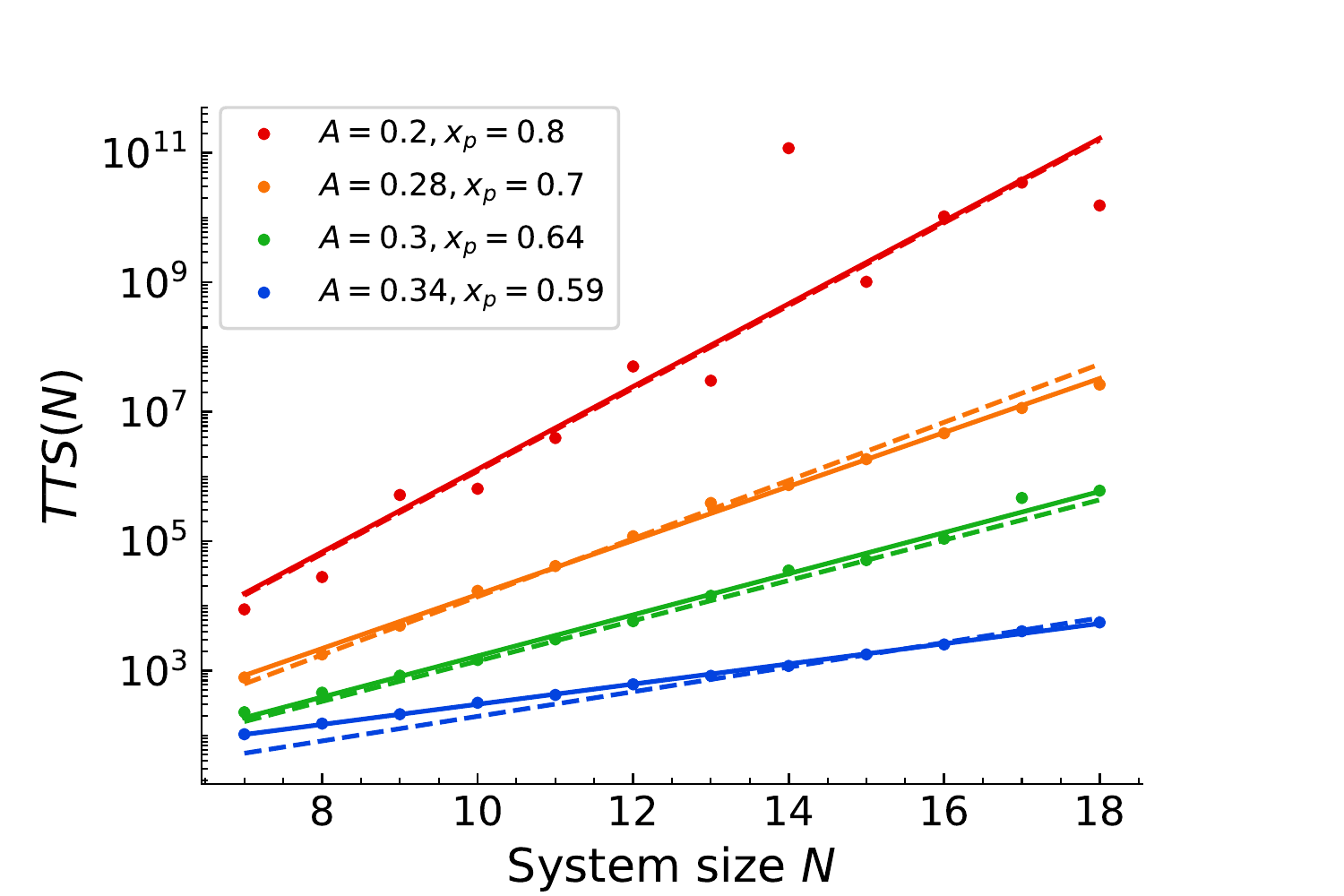}
  \caption{Time to solution to find the true ground state in four problem model sets using the standard uniform sweep method, computed from the final success probability for a runtime polynomially increasing with $N$. The difficulty level of the four models is arranged in descending order: $\{A=0.2, x_p=0.8\}$;  $\{A=0.28, x_p=0.7\}$;  $\{A=0.3, x_p=0.64\}$;  $\{A=0.34, x_p=0.59\}$. Dots are data with the standard uniform sweep method, solid lines are best-fit curves of the numerical data and dashed lines are the inverse of the square of the numerically estimated minimum gap from FIG.~\ref{fig2}. The time to solution closely tracks the average minimum gap squared as expected. In all cases, the $N$-spin tunneling bottleneck in this problem leads to an exponentially increasing time to solution.}
  \label{fig5}
\end{figure}

\begin{comment}
\begin{figure*}
\centering
\subfloat[$A=0.34, x_p=0.59$\centering]{\includegraphics[width = 0.5\textwidth]{standard_0.59_0.34.pdf}\label{fig:f1}}
\hfill
\subfloat[$A=0.3, x_p=0.64$\centering]{\includegraphics[width = 0.5\textwidth]{standard_0.64_0.3.pdf}\label{fig2:f2}}
\hfill
\subfloat[$A=0.28, x_p=0.7$\centering]{\includegraphics[width = 0.5\textwidth]{standard_0.7_0.28.pdf}\label{fig3:f3}}
\hfill
\subfloat[$A=0.2, x_p=0.8$\centering]{\includegraphics[width = 0.5\textwidth]{standard_0.8_0.2.pdf}\label{fig4:f4}}
\caption{Standard uniform sweep routine of four sets of problem model, the difficulty level of the four models is arranged in descending order: $\{A=0.2, x_p=0.8\}$;  $\{A=0.28, x_p=0.7\}$;  $\{A=0.3, x_p=0.64\}$;  $\{A=0.34, x_p=0.59\}$. In each problem set, the time needed to find the solution is getting longer with the increasing system size. And the performance of the standard uniform sweep method is getting worse as the problem getting harder, for example in the $\{A=0.34, x_p=0.59\}$ set, the system can find the true ground state at a very short time in small system size $N=7$ especially compared with the hardest set $\{A=0.2, x_p=0.8\}$.}
\end{figure*}
\end{comment}

We evaluate the performance of the standard uniform sweep algorithm by computing the time to solution($TTS$) over a range of system sizes. The $TTS$ measures the time needed to find the ground state with $99\%$ success probability~\cite{Albash2} 
\begin{equation}\label{TTS}
TTS \propto t_f \frac{\ln(1-0.99)}{\ln(1-p(t_f))},
\end{equation}
where $p(t_f)$ is the success probability in a single-trial with runtime $t_f$; for small $p(t_f)$, $TTS \propto t_f/p(t_f)$. As mentioned in the introduction, the evolution time needed to find the ground state increases as $\Delta_{min}^{-2}$. To explore the performance of the standard uniform sweep method under different problem models, we choose four sets of parameters: $\{A=0.2, x_p=0.8\}$;  $\{A=0.28, x_p=0.7\}$;  $\{A=0.3, x_p=0.64\}$;  $\{A=0.34, x_p=0.59\}$ forming an ensemble of problem models with descending difficulty level. These parameters are chosen to approximately set $\Delta_{min} \propto \left \{ 2^{-N}, 2^{-3N/4}, 2^{-N/2}, 2^{-N/4} \right \}$, respectively. As mentioned previously, increasing $A$ or decreasing $x_p$ toward $1/2$ both decrease the difficulty exponent, and moving either parameter in the opposite direction makes the problem harder.

As expected by an exponentially closing gap, the time needed to find the solution exponentially increases with system size. This is confirmed in FIG.~\ref{fig5}, where the corresponding time to solution exponentially increases with the system size in all problem sets, and the difficulties of the four sets are well separated from each other. With the minimum gap and performance of the standard uniform sweep method rigorously understood, we now apply other methods from the literature to compare their performance with it and investigate their abilities of providing a quantum speedup.

\begin{table*}
\centering
\begin{tabular}{p{3.2cm}p{1.43cm}p{1.43cm}p{1.43cm}p{2.7cm}p{1.43cm}p{1.43cm}p{1.43cm}p{1.43cm}p{0.6cm}}
 \hline
 \hline
Problem set&$1/\Delta_{min}^2$&$S$&$I$&$C_{F,A,M}$&$M$&$CM$&$SyncM$&$SyncMC$&$D$\\
\hline
 $A$=0.2,$x_p$=0.8 & 2.25 & 2.12 & 0.79 & 1.77, 2.09, 2.50 & 1.48 & 1.31 & 1.28 & 0.86 & 1.56\\
\\
  $A$=0.28,$x_p$=0.7 &1.48& 1.39 & 0.70 & 1.25, 1.23, 1.67  & 0.89 & 0.86 & 0.84 & 0.64 & 1.05\\
\\
  $A$=0.3,$x_p$=0.64 &1.04& 1.06 & 0.69 & 0.87, 0.77, 1.12 & 0.62 & 0.62 & 0.64 & 0.62 & 0.50\\
\\
  $A$=0.34,$x_p$=0.59 &0.61& 0.52 & 0.69 & 0.48, 0.44, 0.54 & 0.45 & 0.45 & 0.45 & 0.51 & 0.40\\
 \hline
\end{tabular}

\caption{Summary of scaling exponents for each method. Fitting the time to solution $TTS(N)$ of each method to $2^{\beta+\gamma N}$, the table lists the exponential scaling coefficient ``$\gamma$" value for each method. ``$S$" represents the standard uniform sweep method, ``$I$" represents the Inhomogeneous driving method, ``$C_{F,A,M}$" represents the transverse couplers method, in which ``F,A,M" corresponds to ferromagnetic, anti-ferromagnetic, mixed couplers, respectively. ``$M$" represents the RFQA-M method, ``$CM$", RFQA-M with couplers method, ``$SyncM$", synchronized RFQA-M method, ``$SyncMC$", synchronized RFQA-M with couplers method, and ``$D$" represents the RQFA-D method. }
\end{table*}

\section{summary of results}
Before we proceed to the detailed investigation of alternative QA methods, we compile a summary in TABLE.1, that lists the exponential fitting results of $TTS$ for each method. We fit $TTS(N)$ to $2^{\beta+\gamma N}$ and extract the ``$\gamma$" value to determine the difficulty scaling for each method. We find that in the harder problem sets where ${x_p=0.8,A=0.2}$ and ${x_p=0.7,A=0.28}$, synchronized RFQA-M (with transverse couplers) and Inhomogeneous driving show the best performance, but in the relatively easier problem sets where ${x_p=0.64,A=0.3}$ and ${x_p=0.59,A=0.34}$, the RFQA-D method shows the best scaling advantage. The details are illustrated and discussed in the following sections; we include this table as a central reference point for the results of all of our studies.

\section{\label{sec:DataAnalysis}Modified Adiabatic annealing strategies: inhomogenous driving and transverse couplers}
A wide range of modifications to quantum annealing have shown significant promise in theoretical studies~\cite{Seki,Elizabeth,Hormozi,SelsE3909,vinci2017non,Marshall2018,susa2018quantum,Gra,Hauke_2020}. In this section, we begin applying methods from the literature to our AMP model and assess their performance by computing the $TTS$ as in Eq.~\ref{TTS}. We begin by considering inhomogeneous driving method and transverse couplers. Inhomogeneous driving and the ferromagnetic transverse couplers belong to the class of stoquastic Hamiltonians, while the anti-ferromagnetic couplers and mixed-sign couplers have non-stoquastic Hamiltonians. A stoquastic Hamiltonian has real and non-positive off-diagonal matrix elements in the computational basis~\cite{Sergey}, and can often (but not always ~\cite{Evgeny}) be efficiently simulated by sign-problem-free quantum Monte Carlo (QMC). Non-stoquastic Hamiltonians, on the other hand, suffer from a sign problem and thus cannot be efficiently simulated in QMC in general, though some particular non-stoquastic Hamiltonians can be simulated in QMC by clever schemes to avoid the sign problem~\cite{Ohzeki2017}. Amenability (or not) to QMC is a critical issue in QA, as in recent studies, QMC displays comparable exponential scaling to the physical incoherent tunneling rate in quantum annealers~\cite{Denchev, Isakov, Jiang, Evgeny, Mazzola}. It's thus intuitive to infer that the efficiency of QMC and quantum annealers are similar in solving many problems, making it difficult to realize a genuine quantum speedup. Non-stoquastic Hamiltonians do not suffer from this issue, and have demonstrated significant benefits in some theoretical work~\cite{Seki,Seki_2015,Nishimori,susa2017relation,Hormozi}.

\subsection{Inhomogeneous driving}
In inhomogeneous driving, the transverse fields are ramped down at different rates from one site to the next, as first described in~\cite{Susa2018a,susa2018quantum}. In the original proposal ~\cite{Susa2018a}, the magnitude of the transverse field applied to the $N$ spins is turned off sequentially with a set of time-dependent amplitudes $\Gamma_i(s)$.
In that work, the inhomogeneous driving transverse field circumvents the first-order quantum phase transition and provides an exponential quantum speedup in a p-body interacting mean-field-type model. A more careful analysis \cite{susa2018quantum} which included noise and disorder found the exponential speedup to be somewhat fragile, but showed that a consistent polynomial speedup persisted given these more realistic assumptions. Further, there is experimental evidence that inhomogenous driving is effective in real hardware ~\cite{adame2020}.

Inspired by the performance improvements offered by inhomogeneous driving of the transverse field Hamiltonian, we apply it to the four AMP problem sets as follows
\begin{equation}\label{eq:12}
  \begin{split}
&H(t) = -\frac{1}{N}\sum_{i=1}^{N}\Gamma_i(s)\sigma_i^x+ s H_p,\\
  &\Gamma_i(s) =
    \begin{cases}
      1 & \text{if $s<s_i$}\\
      N(1-s^r) + (1-i) & \text{if $s_i \leq s \leq s_{i-1}$}\\
      0 & \text{if $s_{i-1}<s$}\\
    \end{cases} 
  \end{split}
\end{equation}

\begin{figure}[h]
\centering
  \includegraphics[width=9.5cm,height=7cm]{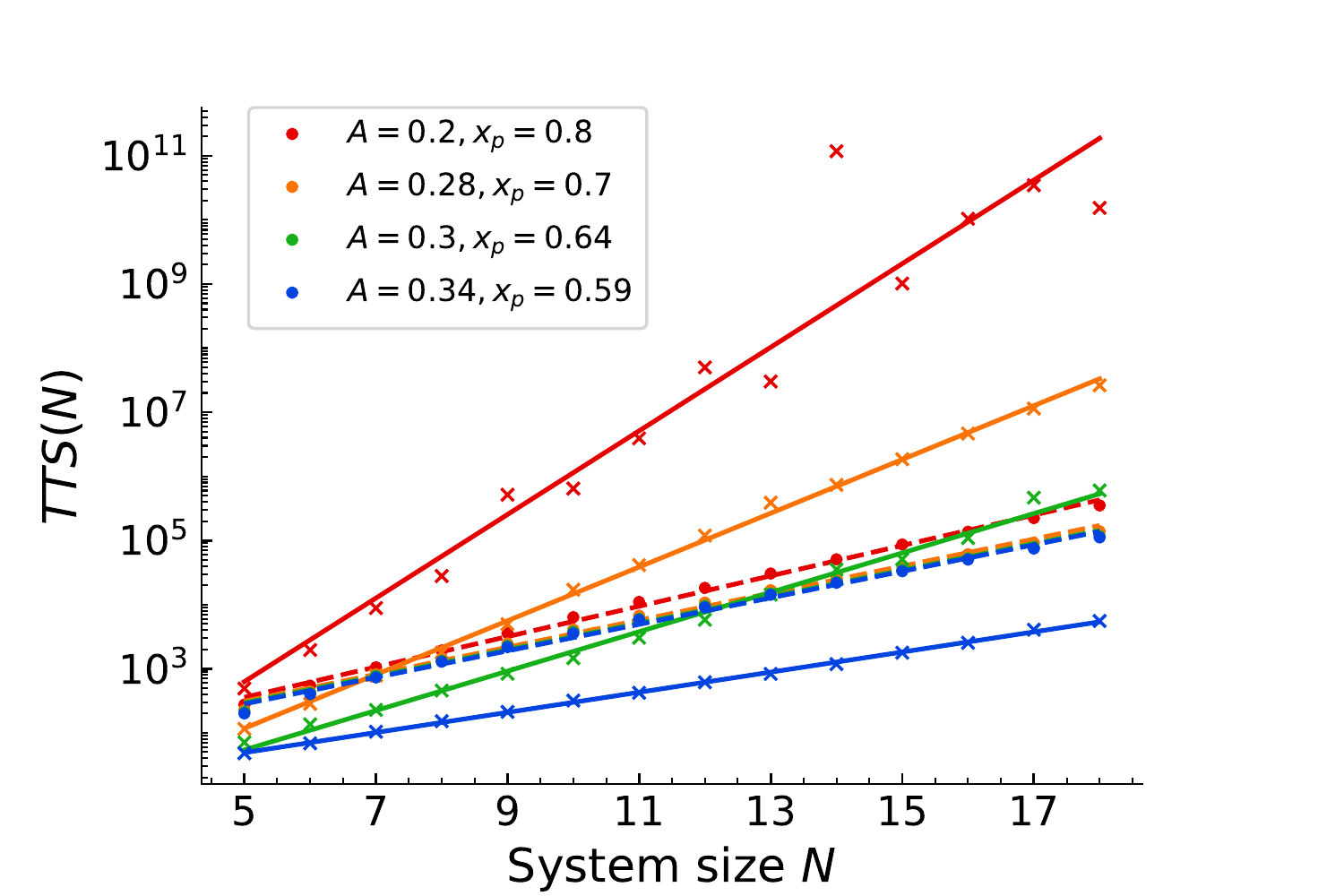}
\caption{Time to find the true ground state in four problem model sets using the inhomogeneous driving method, computed from the final success probability for a runtime polynomially increasing with $N$. Dots are data with the inhomogeneous driving method, cross points are data with the standard uniform sweep  method, dashed lines are best-fit curves of the inhomogeneous driving method, solid lines are best-fit curves of the standard uniform sweep method for comparison purpose. The inhomogeneous driving method helps in the harder cases but in the easier cases it is less efficient than the standard uniform sweep. }
\label{fig6}
\end{figure}

Interestingly, the measurement of $TTS$ in FIG.~\ref{fig6} shows that the inhomogeneous driving method has a difficulty scaling which is very weakly dependent on the control parameters $A$ and $x_p$, with the $TTS$ scaling virtually identically in each case. It consequently outperforms the standard uniform sweep method for the harder problem regimes, but actually shows worse performance for the easiest parameter sets. While we cannot predict its performance analytically in this case (the perturbation theory we use to calculate $\Delta_{min}$ is not well defined for some of the transverse fields set equal to strictly zero), a clue to the origin of this behavior is found in a numerical analysis of the level structure, as we now describe.

\begin{figure}[h]
\centering
  \includegraphics[width=9.5cm,height=7cm]{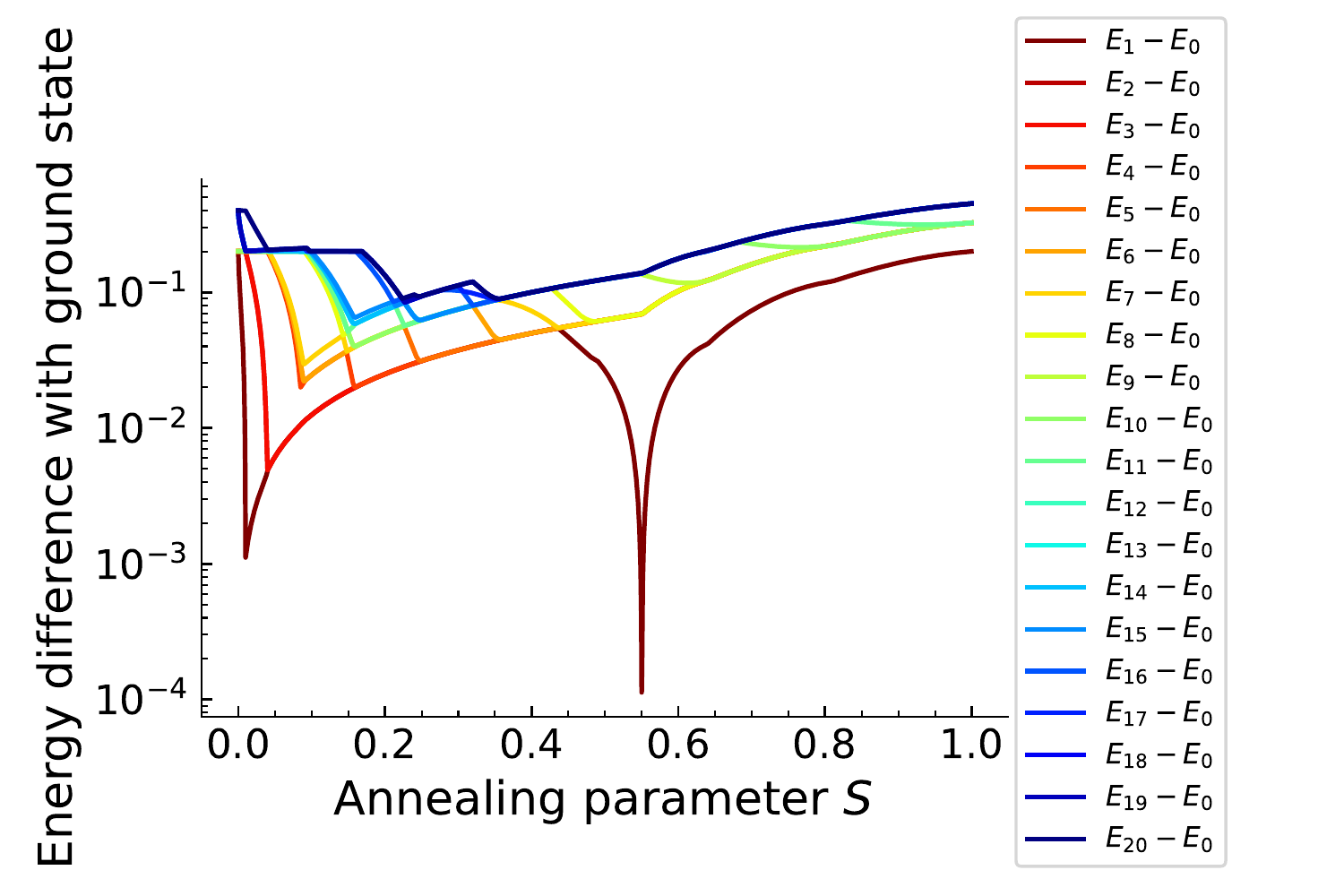}
\caption{Energy difference of the higher order excited states with the ground state. This is the hardest problem where $\{A=0.2, x_p=0.8\}$, $N=10$. We simulated the energy differences up to 20th excited state.}
\label{fig7}
\end{figure}

In FIG.~\ref{fig7} we show the energy difference of the higher order excited states with the ground state in the hardest problem class with $N=10$. In contrast to a uniform sweep, we find two avoided crossings in the annealing process, a generic feature of inhomogenous driving in this system that we observed for other parameter sets as well (data not shown). The presence of two crossings is likely what is responsible for the performance boost observed in the harder problems, and why it seems to have the same scaling for different parameters. A similar phenomenon is observed in the glued trees problem~\cite{somma2012quantum}, where constructive interference of diabatically missing two avoided crossings leads to an exponential speedup. However, unlike the glued trees problem, there is no clear separation between the two competing ground states and the higher excited states in the AMP model. It is clear from FIG.~\ref{fig7} that there also exists an overlap region of the higher order excited states with the first excited state. As $A$ and $x_p$ are varied to make the difficulty scaling decrease, the two avoided crossings move closer together, and the distance from higher levels also shrinks and becomes exponentially small. Consequently, this effect does not result in an exponential speedup here, and shows worse performance than a uniform sweep in the easiest cases.

\begin{comment}
\begin{figure*}
\centering
\subfloat[$A=0.34, x_p=0.59$\centering]{\includegraphics[width = 0.5\textwidth]{ramp0.59.pdf}\label{fig:f1}}
\hfill
\subfloat[$A=0.3, x_p=0.64$\centering]{\includegraphics[width = 0.5\textwidth]{ramp0.64.pdf}\label{fig2:f2}}
\hfill
\subfloat[$A=0.28, x_p=0.7$\centering]{\includegraphics[width = 0.5\textwidth]{ramp0.7.pdf}\label{fig3:f3}}
\hfill
\subfloat[$A=0.2, x_p=0.8$\centering]{\includegraphics[width = 0.5\textwidth]{ramp0.8.pdf}\label{fig4:f4}}
\caption{Ratio of success probability of inhomogeneous driving
method and standard uniform sweep method. The in-homogeneous transverse field can provide some quantum speed up over a homogeneous transverse field. It provides more advantage when the problem is harder, we see inhomogeneous driving method obviously outperform standard uniform sweep routine in the hardest set: $\{A=0.2, x_p=0.8\}$}
\end{figure*}
(the probable reason of why it can effectively remove the phase transition is that a phase transition is a cooperative phenomenon involving all degrees of freedom, and the inhomogeneity of the field would jeopardize the cooperation between different parts of the system)
\end{comment}

\subsection{Transverse couplers}
Adding two-body transverse coupling to QA~\cite{Seki,Seki_2015,susa2017relation,Hormozi} is often considered to be a promising route to a quantum speedup. For instance, Hormozi \textit{et al}~\cite{Hormozi}, constructed a stoquastic Hamiltonian by inserting ferromagnetically coupled term $H_I^F$ to the traditional Ising model, and a non-stoquastic Hamiltonian by inserting antiferromagnetically coupled term $H_I^A$ or mixed coupled term $H_I^M$ as follows
\begin{equation} \label{eq:13}
  \begin{aligned}
    H_I^F = - \sum_{i<j=1}^{N}\sigma_i^x \sigma_j^x\\        
    H_I^A = + \sum_{i<j=1}^{N}\sigma_i^x \sigma_j^x\\
    H_I^M =  \sum_{i<j=1}^{N}r_{ij}\sigma_i^x \sigma_j^x
  \end{aligned}
\end{equation}
$r_{ij}$ is randomly chosen from $\{-1,1\}$ to include both ferromagnetic and antiferromagnetic cases. In that work, they found that both stoquastic and non-stoquastic Hamiltonians showed an advantage over a uniform transverse field for a class of long-range Ising spin glass problems, with the non-stoquastic methods generally showing better performance. This motivated us to investigate the same method in our AMP model. We add transverse couplers into our model and choose a path of the form\cite{farhi,Hormozi}
\begin{equation}\label{eq:14}
H(s) = (1-s)\frac{1}{N}H_0 +s(1-s) \frac{1}{N}H_I + sH_p.
\end{equation}

\begin{figure}[h]
\centering
  \includegraphics[width=9.5cm,height=7cm]{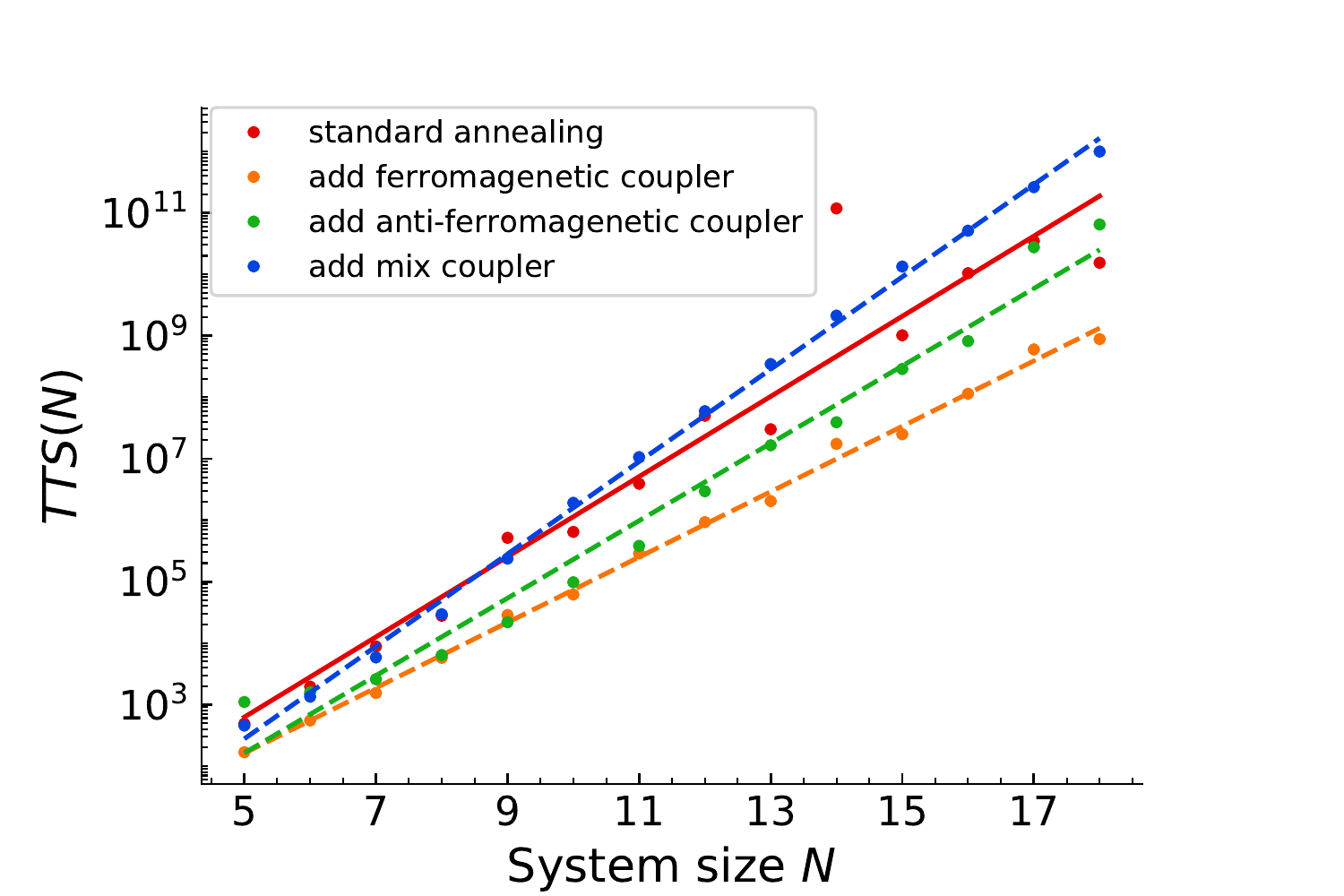}
\caption{Time to find the true ground state in the hardest problem set:$\{A=0.2, x_p=0.8\}$ using the transverse coupler methods, computed from the final success probability for a run-time polynomially increasing with $N$. Data for adding a ferromagnetic coupler, an anti-ferromagnetic coupler, and mixed couplers are given by orange, green, blue dots, respectively. Red dots are data of the standard uniform sweep method for comparison purpose. The solid red line is the best-fit curve of the standard uniform sweep method for comparison. Other dashed lines are best-fit curves for the transverse couplers methods. Adding ferromagnetic and anti-ferromagnetic couplers to the conventional standard uniform sweep routine show obvious quantum speed up, although adding the mixed couplers reduces the advantage to some extent. }
\label{fig8}
\end{figure}

\begin{figure*}
\centering
\subfloat[$A=0.34, x_p=0.59$\centering]{\includegraphics[width = 0.5\textwidth]{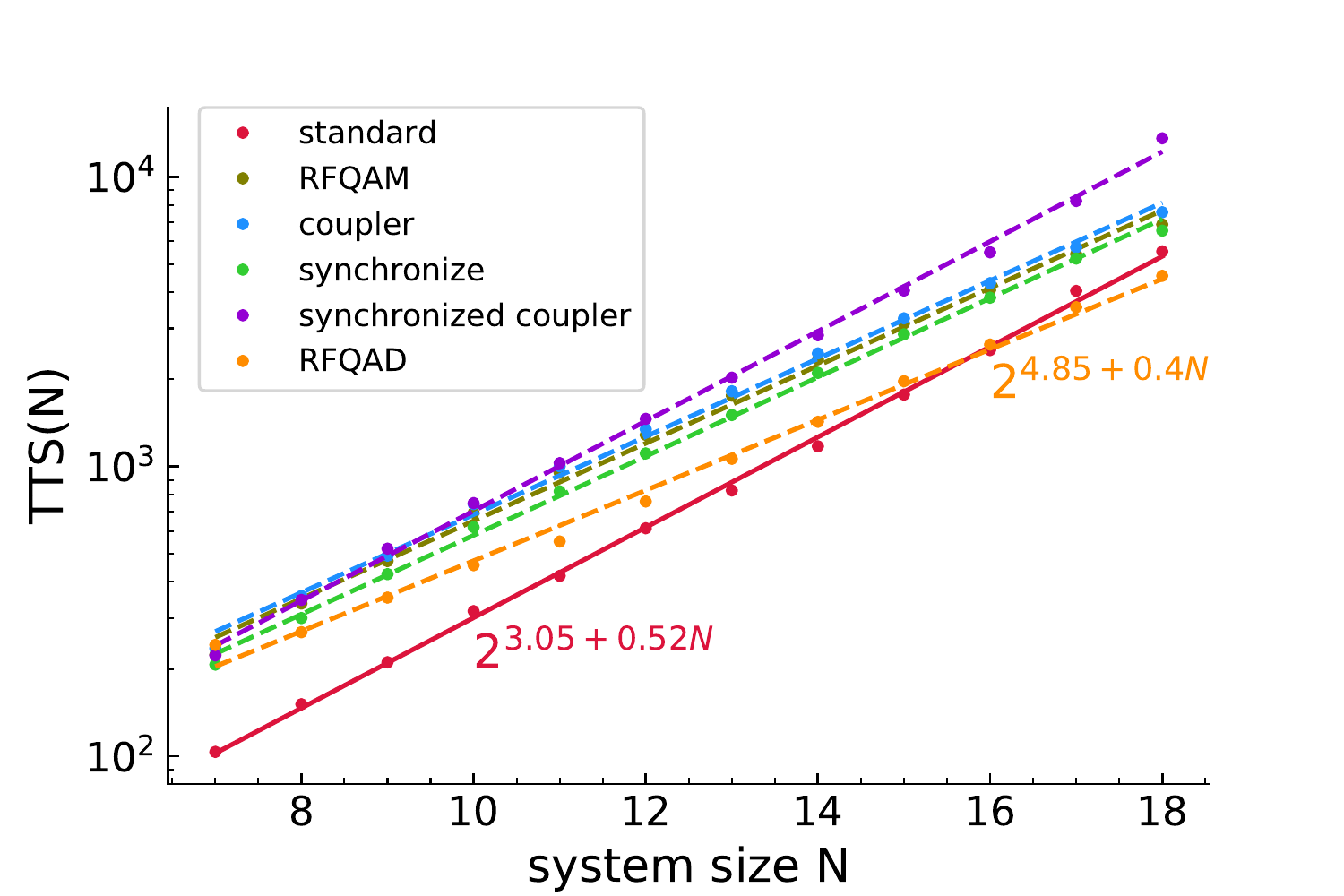}\label{fig:f1}}
\hfill
\subfloat[$A=0.3, x_p=0.64$\centering]{\includegraphics[width = 0.5\textwidth]{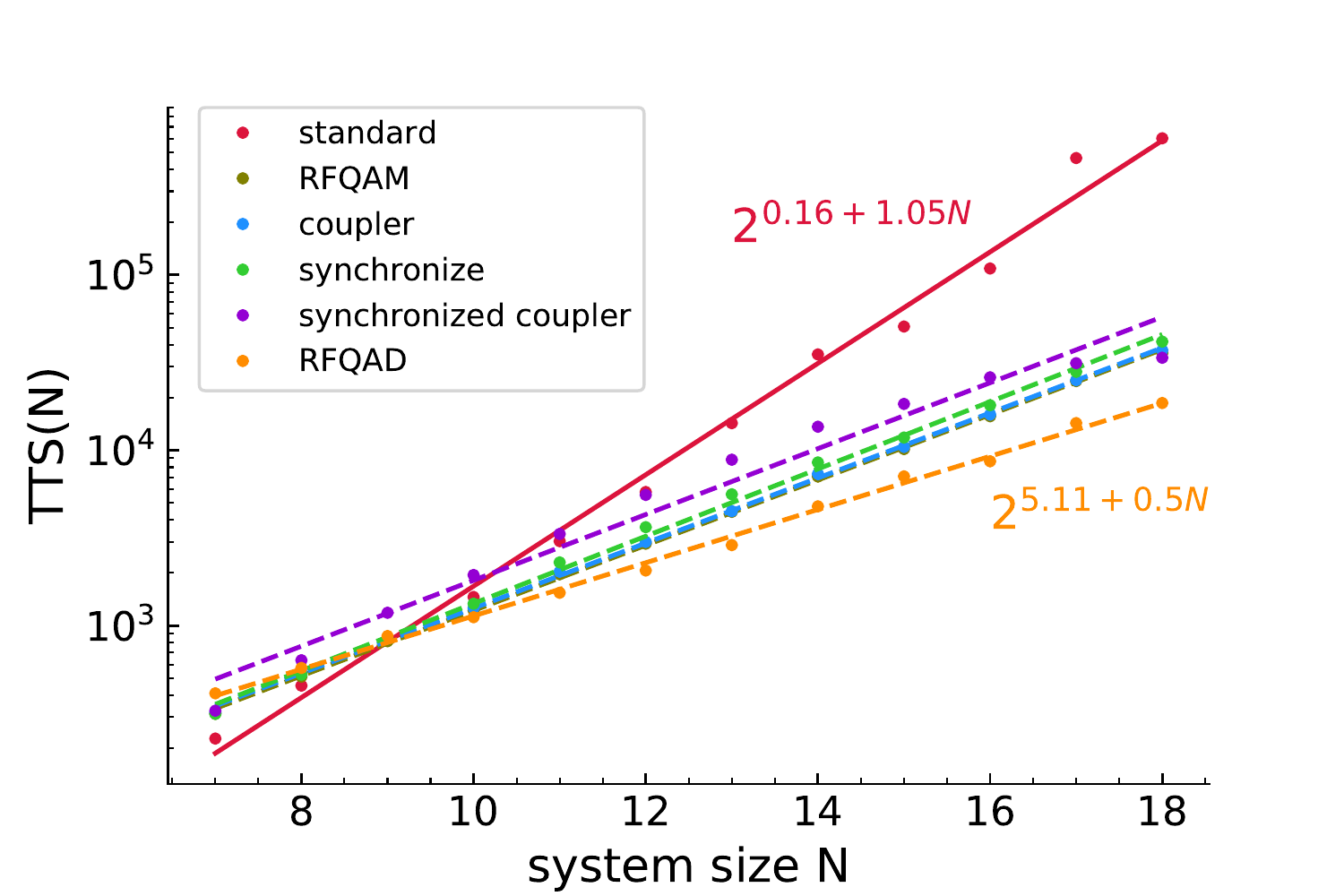}\label{fig2:f2}}
\hfill
\subfloat[$A=0.28, x_p=0.7$\centering]{\includegraphics[width = 0.5\textwidth]{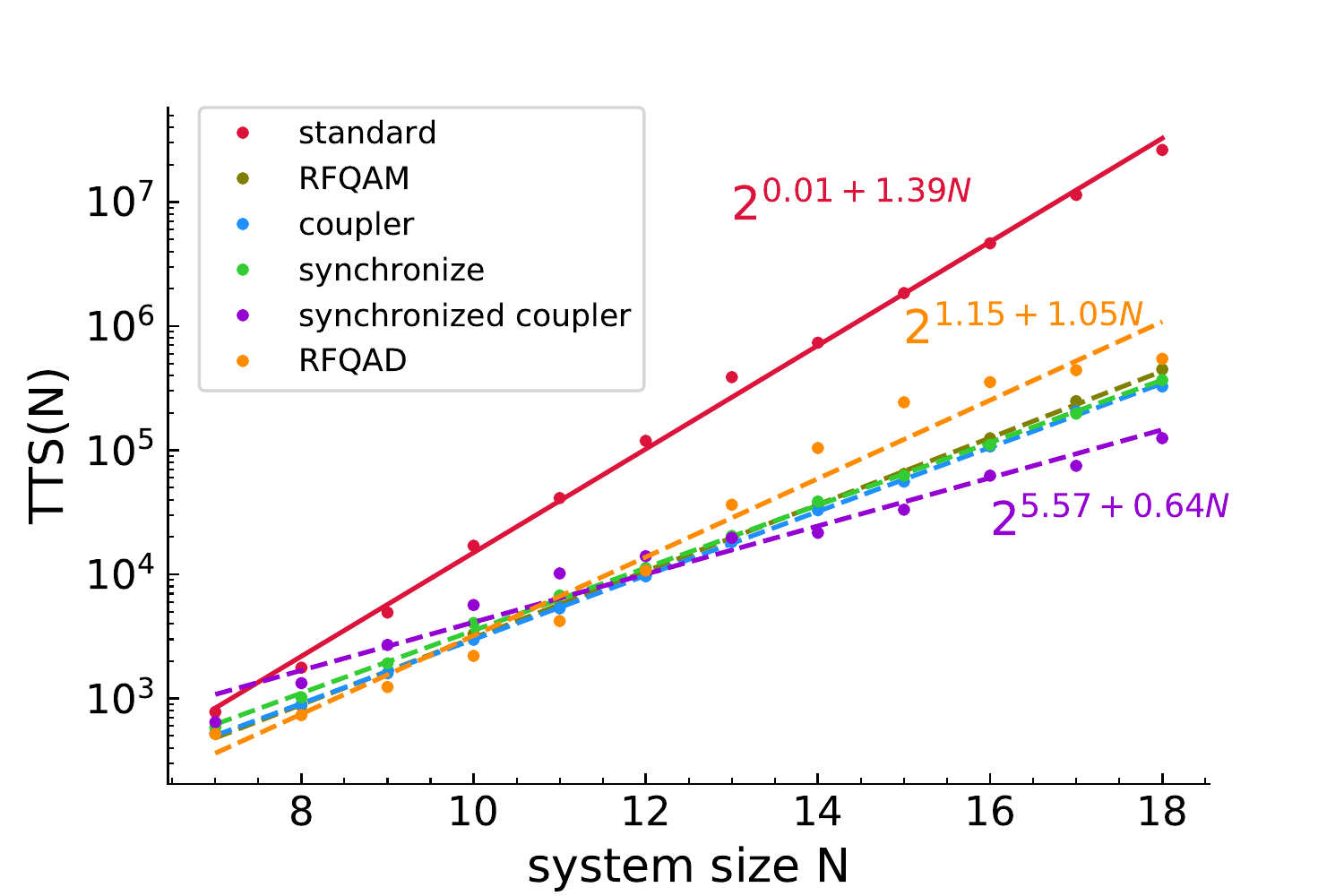}\label{fig3:f3}}
\hfill
\subfloat[$A=0.2, x_p=0.8$\centering]{\includegraphics[width = 0.5\textwidth]{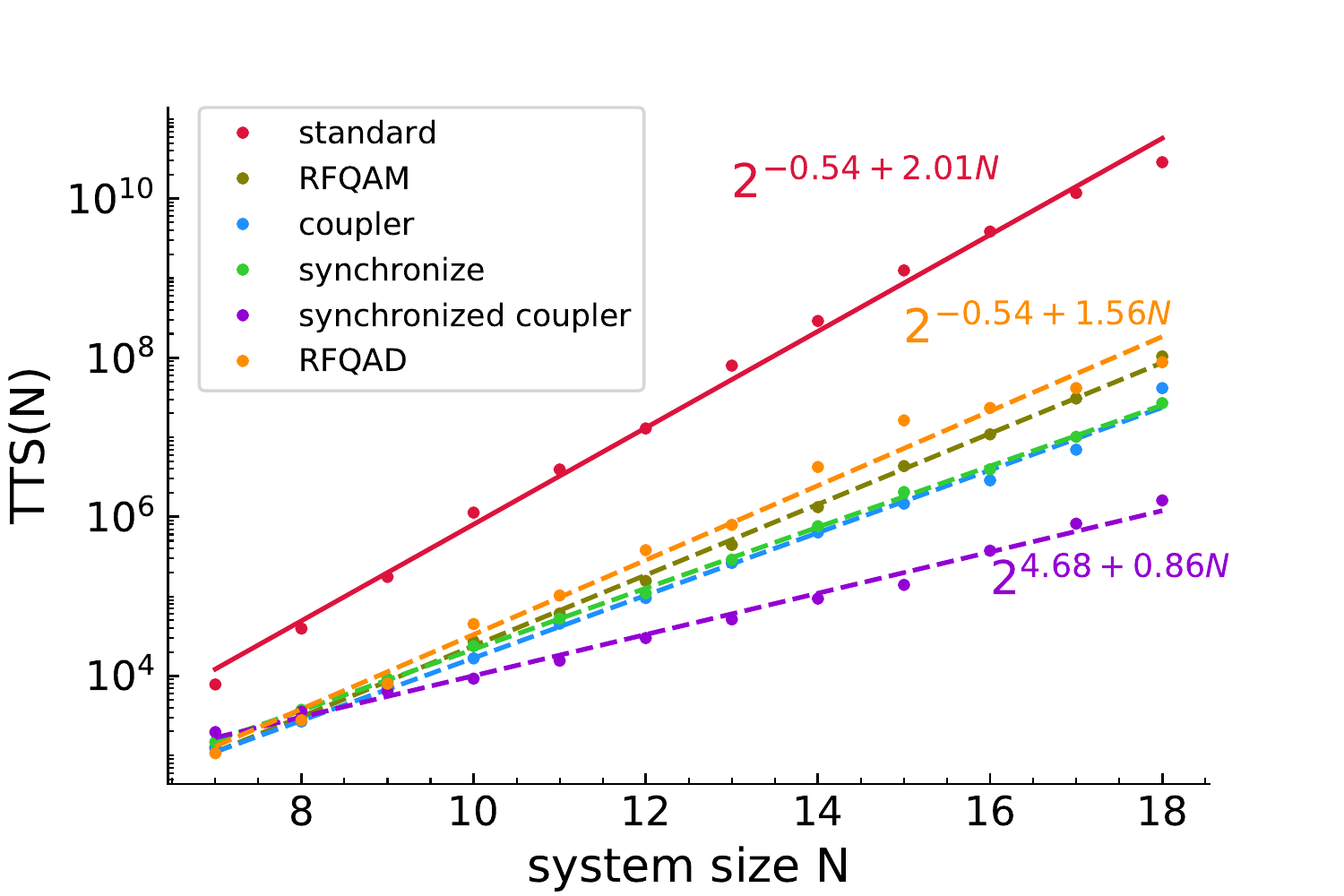}\label{fig4:f4}}
\caption{Time to find the true ground state in four problem model sets using the RFQA method, computed from the final success probability for a run-time polynomially increasing with $N$. Dots are data with RFQA methods, the solid red line is the best-fit curve of the standard uniform sweep method for comparison purpose. Other dashed lines are best-fit curves of the RFQA methods. RFQA methods show obvious scaling advantages over the standard quantum annealing, with generally greater relative improvements in the harder problem cases.}
\label{fig9}
\end{figure*} 

\begin{comment}
stoquastic Hamiltonians have only real and non-positive off-diagonal entries, so that their path-integral configurations have real and non-negative contributions to the partition function.This property of stoquastic Hamiltonian can help it to circumvent the so-called ``sign problem"~\cite{E.Y}, correspondingly the non-stoquastic terms can’t avoid sign problem and thus can’t be efficiently simulated in Quantum Monte Carlo, then it will be easy to simulate a quantum speedup with non-stoquastic terms. (from the literature, coupled driver terms, adding ferro coupler would introduce a stoquastic Hamiltonian, but antiferro or mixed coupler would introduce non-stoquastic Hamiltonian, (stoquastic or non-stoquastic or mixed ) could outperform the standard uniform sweep routine,)(Nonstoquastic Hamiltonians greatly outperform their stoquastic counterparts and their superiority persists as the system size grows, the improved performance is probably related to the frustrated nature of nonstoquastic hamiltonians.)
\end{comment}
We apply the transverse coupler Hamiltonian to our four problem sets, plotting the results for the hardest scaling choice as an example in FIG.~\ref{fig8}. It is straightforward to see that adding ferromagnetic or anti-ferromagnetic coupler has a clear scaling advantage over a standard uniform sweep, but mixed couplers actually lead to decreased performance. The quantum speedup from coupler terms is probably because the couplers can flip two spins simultaneously, so the tunneling process from one configuration to the other can occur at lower order than with a uniform transverse field (where it occurs at Nth order in this model). The ferromagnetic coupler increases the minimum gap and thus provides a quantum speedup over the standard uniform sweep method, the same effect is observed in\cite{Hormozi}. The antiferromagnetic couplers actually decreased the minimum gap but still show a scaling advantage, so the reason for the increased performance from the antiferromagnetic ones remains elusive. The behavior of the transverse coupler methods in other three problems sets are shown in TABLE I. 

% The reason of why adding coupler method can outperform the standard uniform sweep method is probably the coupler terms would contribute additional sources to the proliferation of multi-photon transitions, which provides performance increases~\cite{Hormozi,Kapit2017}.

\section{Reverse annealing and cold baths}

Reverse annealing~\cite{perdomo2011study,chancellor2017modernizing,ohkuwa2018reverse,king2018observation}, where the system is initialized in a local classical minimum and the transverse field is ramped up and down to search for other minima, unfortunately provides no benefit for the AMP. Reverse annealing was shown in~\cite{ohkuwa2018reverse} to provide benefit for the p-spin ferromagnet problem, if one is able to guess an initial state sufficiently close to the true ground state. However, in the AMP there are only two minima to choose from, separated by $N$ spin flips. The only sensible choice (without dramatically modifying $H_P$) is thus to initialize the system in the false minimum. We simulated the reverse annealing protocol (data not shown) by initializing the system in the false minimum, ramping the transverse field up to a finite value guessed randomly from an $O(1)$ range enclosing the phase transition point, evolving from that point for $O \of{N^2}$ time, then ramping it back down to zero. With sufficient averaging over the location of the pause point (which is not knowable precisely in real problems), we found a time to solution which scaled nearly identically to the standard uniform sweep method for all parameters studied. Thus, we found no benefit in applying reverse annealing to this problem. 

The influence of a cold bath on this system is more subtle. It is well known~\cite{keck2017dissipation,smelyanskiy2017quantum,venuti2017relaxation,arceci2018optimal,marshall2019power,kadowaki2019experimental,suzuki2019quantum,roberts2019noise} that coupling a quantum spin glass to a cold bath can improve the process of finding its low energy states. So let us consider coupling the AMP to a low temperature bath during annealing. Importantly, we here assume that $T$ is small compared to the single qubit excitation energy, but it may still be large compared to the (exponentially small) minimum gap. How much can such a bath improve the time to solution?

Unfortunately, numerical simulation of such a system is prohibitively expensive~\cite{jaschke2019thermalization} given the complexity of the Lindblad operators used to represent the finite temperature bath. We can however estimate the relaxation rate from the bath by appealing to the MSCALE conjecture~\cite{Kapit2017}. This conjecture states that, for few-body operators, the scaling (with problem size $N$) of matrix elements of these operators between competing ground states of quantum spin glasses near a phase transition is the same as the scaling of the minimum gap itself. This conjecture is true by inspection for the AMP, since the gap can be computed accurately using the modified forward approximation in Sec.~\ref{sec:Materials}. If we assume that each spin couples to a cold bath independently, then the rate of mixing near the phase transition scales as $N \Gamma_B^2 /W $ , where $\Gamma_B \propto \Delta_{\rm min}$ is the matrix element from a local spin operator and $W$ is the energy range swept over. This produces a factor of $N$ enhancement relative to the closed system, but does not change the scaling exponent as the other methods do. The cold bath may however improve performance in a real system by relaxing few-body excitations back toward the ground state, correcting ``errors" induced by other channels.

\section{RFQA}

%The above methods are inspired by the performance of adding inhomogeneous transverse field~\cite{susa2018quantum} and the potential of a non-uniform drive Hamiltonian~\cite{Lanting2017}. 

% We introduce a new method in this section, which cannot be simulated in QMC and we provide theoretical prediction of its potential to provide a quantum speedup. We name the method RFQA (RFQA stands for random field quantum annealing)~\cite{Kapit2017}. 

Stoquastic or not, the previous sections all explored ``DC" schemes involving slow variations of transverse field and coupler terms. In this section, we consider an AC alternative, called RFQA~\cite{Kapit2017}. In RFQA, the traditional transverse field driver Hamiltonian is modified by  independently oscillating either the magnitude (RFQA-M) or direction (RFQA-D) of each transverse field term (M and D refer to magnitude and direction, respectively). As we will describe shortly, the qualitative explanation for a quantum speedup in RFQA is an exponential proliferation of weak many-spin processes, leading to accelerated mixing near first order quantum phase transitions. The total Hamiltonian in RFQA is given by
\begin{equation}\label{eq:15}
H(t) = (1-s)H_{M/D}(t) + sH_p,
\end{equation}
where the driving fields in RFQA-M and RFQA-D are defined as follows
\begin{equation}\label{eq:16}
  \begin{split}
    &H_M(t) = -\kappa \sum_{i=1}^{N}(1+ \bar{\alpha_i} \sin(2\pi f_i t))\sigma_i^x,\\
    &H_D(t) = -\kappa \sum_{i=1}^{N} [\cos(\bar{\alpha_i}\sin(2\pi f_i t))\sigma_i^x +\\
    &\ \ \ \ \ \ \ \ \ \ \ \  \sin(\bar{\alpha_i}\sin(2\pi f_i t))\sigma_i^y].
  \end{split}
\end{equation}
Here, $\bar{\alpha_i}$ is the amplitude of each oscillation, the frequencies $f_i$ of the field are randomly chosen between $f_{min}$ and $f_{max}$, and $\kappa$ is the magnitude of the transverse field. To avoid uncontrolled heating, both $f_{min}$ and $f_{max}$ have inverse polynomial scaling in N. To estimate the performance of RFQA, we average the success probability $p(t_f)$ over hundreds of random choices of the $\{ f_i \}$ when computing time to solution. The RFQA methods all rely on finite frequency dynamics that are not captured by QMC, making them promising candidates for producing a quantum speedup. The two methods are straightforward to implement in flux qubit hardware, by applying oscillating magnetic fields as described in~\cite{Kapit2017}.

As described in the original work, the qualitative speedup mechanism from RFQA is complex and arises from an exponential proliferation of weak multi-photon transitions. As the system nears a phase transition point, whenever the energy of the two ground states crosses a combination of $m$ oscillating frequencies there is an $m$th order driving process that (very weakly) mixes the two states. In general, the Rabi frequency of such a process decreases exponentially in $m$, but there are $2^m \binom{N}{m}$ such terms and the combination of all of them dramatically accelerates the phase transition. If the $m$th order resonance is smaller than the base tunneling rate $\Omega_0 = \Delta_{min}/2$ by a factor $\Lambda^m$, then the total transition rate is expected to scale approximately as
\begin{eqnarray}\label{eq:20}
\Gamma_T &=& \frac{\sum_{i=1}^{N} |\Omega_i|^2}{W} \simeq \frac{\Omega_0^2}{W} \sum_{l=1}^{N} \Lambda^{2l}{N \choose l}2^l \\
&\simeq& \frac{\Omega_0^2}{W} \left (1+2 \Lambda^2 \right)^N.  \nonumber
\end{eqnarray}
Predicting $\Lambda$ is a subtle challenge and something we will leave for future work; we restrict our study of RFQA to purely numerical simulations here.

\subsection{RFQA-M}
% Based on RFQA-M and adiabatic annealing strategies, we make some variations of RFQA-M to make further investigation on RFQA-M:
%(1)Partially synchronize RFQA-M method, in which $N$ spins are broken into $k$ groups, and the transverse fields in each group are all oscillated in phase with the same frequencies. In our work, we only divide the $N$ spins into 2 groups for now.
%(2)Adding ferromagnetic/antiferromagnetic couplers in the RFQA-M method, this method integrates the promising add transverse coupler method with RFQA-M, while the added transverse couplers are oscillated in magnitude, the total Hamiltonian in this method is defined as follows.

In RFQA-M the magnitudes of the transverse field terms coherently oscillate with time as the global amplitude is ramped down toward 0, so that an individual transverse field term $\kappa$ is replaced with $\kappa \left(1 + \bar{\alpha} \sin 2 \pi f_i t \right)$. In our simulations we used $\bar{\alpha}=0.9$, and magnitude of frequencies $f_i$ is randomly chosen between $\{\frac{0.01}{N^{1.5}}, \frac{0.02}{N{1.5}}\}$; the signs of the $f_i$ are also randomly chosen. This is superficially similar to inhomogenous driving, but the coherent oscillations lead to non-monotonic changes of $\kappa_i$ with time and very different scaling as a result. We also considered a few additional variations of RFQA-M. In one set of simulations, we explored a partially synchronized RFQA-M method, in which $N$ spins are broken into $k$ groups, instead of generating different random frequencies for each site, the transverse fields in each group are all oscillated in phase with the same frequencies. In this work, we only divided the $N$ spins into two groups, but other arrangements are possible. We also explored adding ferromagnetic/antiferromagnetic couplers to the RFQA-M method, where all transverse couplers and fields are independently oscillated in magnitude. The total Hamiltonian in this method is defined as follows

\begin{equation}\label{eq:19}
\begin{aligned}
H(s) &= (1-s)  H_{M/D}+ sH_p \\ & + (1-s)\kappa_r\sum_{<i,j>}\sin(2\pi r_{ij}t)\sigma_i^x\sigma_j^x
\end{aligned}
\end{equation}
where $\kappa_r$ is the magnitude of the coupler terms, $r_{ij}$ are the oscillating frequencies of the coupler, randomly chosen between $\{r_{min},r_{max}\}$. The magnitude $r$ is defined to polynomially decrease with $N$, and the frequencies $r_{ij}$ are also inverse polynomial in $N$. Finally, we looked at partially synchronized RFQA-M with transverse couplers, where the transverse couplers are also synchronized into groups. 

We compare the $TTS$ of the various implementations of RFQA-M with the standard uniform sweep method in FIG.~\ref{fig9}. The results show that RFQA-M and its adaptions can provide quantum speed up over the standard uniform sweep routine, with a scaling advantage which is particularly obvious in harder problem sets.
\subsection{RFQA-D}
% In RFQA-D, the oscillating direction of the transverse field preserves the energy of the system. To analyze the performance of the RFQA-D method, we sum up the contributions from multi-photon resonance to make a prediction of solution rate of RFQA-D

% Based on the analysis of minimum gap, we fit the solution rate of RFQA-D method in the hardest problem model $\{A=0.2,x_p = 0.8\}$ which goes as $2^{1.4-0.96N}$. The standard uniform sweep routine applied to the problem set has a solution rate of $2^{0.54-2.01N}$, which represents RFQA-D is supposed to outperform the standard annealing routine. As shown in Fig.~\ref{fig9}, we see RFQA-D does provide obvious quantum speed up. RFQA methods show the most promise for improving the TTS of our trial problem, as their performance scales favorably with the difficulty exponent of the problem. We expect that these results carry over to the larger class of optimization problems that experience wrong-way steering towards false minima.

In RFQA-D the direction of each transverse field term oscillates in time, tipping back and forth in the $x-y$ plane. This can be engineered through oscillating $z$ biases (which can be shown to be equivalent to a tipping transverse field through a time-dependent unitary transformation) and has the elegant property that the instantaneous spectrum of the system is preserved in the evolution, so the oscillations have no ``steering" effect whatsoever (unlike RFQA-M, where changing transverse field magnitudes can change the relative energies of competing ground states, in addition to any AC effects). Any performance advantage from RFQA thus comes directly from the proliferation of weak transitions described above.

As shown in FIG.~\ref{fig9}, we see that RFQA-D does provide an obvious quantum speed up over a uniform sweep. In all studied cases, RFQA-D reduced the exponent for $TTS(N)$ relative to the standard uniform sweep, and for the two easier difficulty regimes, it outperformed all other studied methods. We expect that these results will carry over to the larger class of optimization problems that experience wrong-way steering towards false minima.

\section{Conclusion}
In this work, we defined a simple toy model-- the asymmetric magnetization problem-- with two competing ground states separated by a global peak, and used it to benchmark a variety of modifications to quantum annealing in the literature. The problem is exponentially difficult to solve due to its exponentially closing gap, and the entropic steering toward a false minimum responsible for its difficulty is a generic bottleneck mechanism for a huge array of optimization problem classes. Thus, methods to accelerate finding the solution in it should prove beneficial in much broader contexts.

We studied an ensemble of problem model sets with descending difficulty: $\{A=0.2,x_p = 0.8\}; \{A=0.28,x_p = 0.7\}; \{A=0.3,x_p = 0.64\}; \{A=0.34,x_p = 0.59\}$, and assessed a variety of new quantum methods by evaluating the scaling of the time to solution ($TTS$). To have a straightforward view of the performance of each method, we fit their $TTS$ to exponential functions and extracted the exponential scaling value, summarized in TABLE I. The standard uniform sweep method shows inverse gap squared dependence as expected. In contrast, in the inhomogeneous driving approach, the $TTS$ has a nearly constant difficulty scaling of $2^{\sim 7 N/10}$, roughly independent of the tuning parameters. It likely means that the problem is steered less toward the false minimum in this case than it is for uniform driving, but that is not enough to avoid a first order transition and the resulting lack of guidance becomes counterproductive when the problem is easier. 

While the problem we studied is homogeneous (in that energy is a function of total magnetization only), we don’t expect disorder to significantly change the results for the standard uniform sweep, couplers and RFQA. If we modeled disorder as a simple random $Z$ term with magnitude ~$1/N$ for each spin (recall that the problem energy is $O \left( 1\right)$) added to $H_p$, it would change the relative energies of the ground states by $1/\sqrt{N}$ for the energy scale we have chosen. This  will move the transition point $s_c$ around from one instance to the next, but by an amount that vanishes as $N \rightarrow \infty$. Given that our modified forward approximation calculation predicts $\Delta_{min}$ fairly accurately, examination of those equations shows that this change should not effect the scaling exponent at large $N$. However, it might effect inhomogeneous driving more significantly, as has been seen in other problems~\cite{somma2012quantum,Susa2018a,susa2018quantum,crosson2020}. 

For the transverse coupler method, both ferromagnetic or anti-ferromagnetic coupler terms can provide obvious improvements, but adding a mixture of ferromagnetic and anti-ferromagnetic coupler terms proved counterproductive. We expect that the speedup from the coupler terms arises from creating more tunneling paths between the two competing ground states, since each coupler flips two spins simultaneously. Interestingly, we saw very similar scaling benefits for both ferromagnetic (stoquastic) and anti-ferromagnetic (non-stoquastic) coupler methods; there is no obvious connection between non-stoquasticity and increased performance in this problem. 

Among the RFQA methods, synchronized RFQA-M with the added couplers provided the greatest quantum speedup in the two hardest problem sets, while RFQA-D showed the best scaling in the easier problem instances. The speedup mechanism for both couplers and RFQA methods is due to an amplification of the tunneling rate and has nothing to do with local energetic guidance. 

We conclude that although we did not achieve an exponential speedup for this problem, all of the methods can provide a quantum speedup over the standard uniform sweep routine: inhomogeneous driving provides a significant boost in harder problem sets; transverse couplers added to the standard uniform sweep routine create more tunneling paths between two competing ground states and help decrease the time needed to find the solution; and the introduction of oscillating fields in the RFQA methods can help to stimulate multi-tone transitions, providing more possibilities for the two competing ground states to mix. Given that all three of inhomogeneous driving, RFQA-M and RFQA-D only require modification to the control circuitry and not the qubit hardware itself, we see them as the most promising and cost-effective routes to a near-term benefit. It would be worthy to continue to study these methods on realistic problems at larger scales, and it requires minimal changes to existing hardware to verify their potential in experiment.
\\
\section*{Acknowledgements}
ZJ.T thanks Nicholas Materise and David Rodriguez Perez for helpful discussions. This work was supported by the National Science Foundation through grant number: PHY-1653820. It was made possible by the high performance computing resources from the Tulane University Cypress platform,and the Colorado School of Mines Wendian platform.

\bibliographystyle{unsrt}
\bibliography{bibfile.bib}

\begin{comment}
\section{\label{sec:Bibliography} Reference}
\begin{itemize}
    \item $[1]$ author, \textit{title}, university (, place, year).
\end{itemize}
\end{comment}

\end{document}